\documentclass{ws-mpla}
\usepackage{cite}

\begin{document}
           
\markboth{S. Alekhin, J. Bl\"{u}mlein, S.-O. Moch}
{Determination of $\alpha_s$ and $m_c$ in DIS}

\catchline{}{}{}{}{}

\title{{\footnotesize \vspace{-2.775cm} 
    ~~~~~~~~~~~~~~~~~~~~~~~~~~~~~~~~~~~
    DESY 13-121, DO-TH 13/17, SFB/CPP-13-44, LPN13-041 \vspace{2.775cm}} \\ 
DETERMINATION OF 
$\alpha_s$ 
AND $m_c$ IN DEEP-INELASTIC SCATTERING}

\author{\footnotesize SERGEY ALEKHIN\footnote{
Permanent address: Institute for High Energy Physics, Pobeda 1, 
Protvino, 142280, Russia}}
\address{Deutsches Elektronen-Synchrotron DESY, Platanenallee 6\\
Zeuthen, D–15738, Germany\\
sergey.alekhin@desy.de}

\author{JOHANNES BL\"{U}MLEIN}

\address{Deutsches Elektronen-Synchrotron DESY, Platanenallee 6\\
Zeuthen, D–15738, Germany\\
johannes.bluemlein@desy.de}

\author{SVEN-OLAF MOCH}

\address{II. Institut f\"ur Theoretische Physik, Universit\"at Hamburg, Luruper
  Chaussee 149\\  Hamburg, D-22761, Germany\\
sven-olaf.moch@desy.de}

\maketitle

\pub{Received (Day Month Year)}{Revised (Day Month Year)}

\begin{abstract}
We describe the determination of the strong coupling constant $\alpha_s(M_Z^2)$ and 
of the charm-quark mass $m_c(m_c)$ in the $\overline{\rm MS}$-scheme, based on the 
QCD analysis of the unpolarized World deep-inelastic scattering data. At NNLO the 
values of $\alpha_s(M_Z^2)=0.1134\pm 0.0011(\text{exp})$ and 
$m_c(m_c)=1.24 \pm 0.03 (\text{exp})\,^{+0.03}_{-0.02}
(\text{scale})\,^{+0.00}_{-0.07} (\text{th})$ are obtained and are compared 
with other determinations, also clarifying discrepancies. 

\end{abstract}

\ccode{PACS Nos.: 12.38.Qk, 12.38.Bx.}

\section{Introduction} 

The process of lepton-nucleon deep-inelastic scattering (DIS)  
is a clean source of basic information about the hadron substructure in terms 
of the parton model. Moreover, the QCD corrections to the parton model 
provide the connection of the DIS structure functions with the parameters 
of the QCD Lagrangian, in particular to the strong coupling $\alpha_s$
and the heavy-quark masses. 
The higher order QCD corrections are manifest in the scaling violations 
of the structure functions w.r.t. the virtual photon momentum transfer $Q^2$. 
This phenomenon was observed shortly after the discovery of the partonic structure 
of the nucleon and provided one of the first constraints on $\alpha_s$. 
With the dramatic improvement in the accuracy of the lepton-nucleon
DIS data and the progress in the theoretical calculations the value of $\alpha_s$ 
can in principle be determined with an accuracy of $O(1\%)$~\footnote{For a recent overview
on precision determinations of $\alpha_s(M_Z^2)$ see \cite{Bethke:2011tr}.}.
The scaling violations, however, are also sensitive to the parton 
distribution functions (PDFs). Therefore the determination of $\alpha_s$
has to be performed simultaneously with the nucleon PDFs in global fits. 
Furthermore, an elaborate theoretical description of the 
QCD scaling violations is available for the leading-twist terms only. 
In practice this requires a careful isolation of the higher-twist effects 
and/or their independent phenomenological parameterization. 
Another important aspect of DIS phenomenology is related to the 
$c$- and $b$-quark contributions. The heavy-quark production cross
section is sensitive to the heavy-quark masses. Therefore the DIS 
data provide a constraint on the $c$- and $b$-quark masses, $m_{c,b}$.  
The structure functions of the semi-inclusive process with the heavy 
quark in the final state are particularly useful for this purpose, although
the data on inclusive structure functions are competitive with the 
semi-inclusive ones due to a much better accuracy. A major pitfall arising
in the analysis of heavy quark production is related to the account of 
the higher-order QCD corrections. Due to the two scales appearing in the problem 
the calculations are quite involved. Therefore the NNLO corrections 
to the heavy-quark lepto-production are known in partial form 
only~\cite{Kawamura:2012cr} at present. The problem of the high-order
corrections is bypassed in the so-called variable-flavor-number (VFN) scheme 
assuming zero mass for the $c$- and $b$-quark. In this approximation 
the available high-order massless DIS Wilson coefficients can be 
employed for the calculation of the heavy-quark lepto-production rates. 
As well-known, the VFN approximation is obviously inapplicable  at scales 
$Q^2 \sim m_{c,b}^2$ and it is commonly supplemented by {\it modeling} of the Wilson 
coefficient at low $Q^2$ in order to arrive in this way at a general-mass VFN (GMVFN) 
scheme. In the present paper we essentially focus on the determination of 
$\alpha_s$ and $m_c$ based on the fixed-flavor-number (FFN) scheme. Here 
the mass effects are taken into account on field-theoretic grounds, 
free of model ambiguities appearing in the so-called GMVFN schemes. Moreover, we employ 
the massive Wilson coefficients derived using the running-mass definition, which provide 
improved perturbative stability of the heavy-quark production rate~\cite{Alekhin:2010sv}.

The paper is organized as follows. 
In Section~\ref{sec:basics} we outline the theoretical basis of the 
analysis and describe the data used. Sections~\ref{sec:alphas} and \ref{sec:mass} contain
our results on the determination of $\alpha_s$ and the $c$-quark mass, 
respectively. In Section~\ref{sec:vfn} we compare the FFN and 
VFN schemes with particular emphasis on the uncertainties 
in the determination of $\alpha_s$ and $m_c$. 

\section{Theoretical and Experimental Ingredients of the Analysis}
\label{sec:basics}

Our determination of $\alpha_s$ and $m_c$ is based on the QCD analysis 
of the DIS data obtained in fixed-target experiments and at the HERA
collider.
Only the proton- and deuteron-target samples are selected which
allows to minimize the impact of the nuclear corrections on the results~\footnote{For a 
detailed description 
of the data set used cf. Ref.~\cite{Alekhin:2012ig}.}. 
 The DIS
data are combined with the ones on the fixed-target Drell-Yan process, 
providing a supplementary constraint on the PDFs and 
to facilitate the separation of the valence and sea quark distributions.  
%
\begin{figure}[pt]
\centerline{\psfig{file=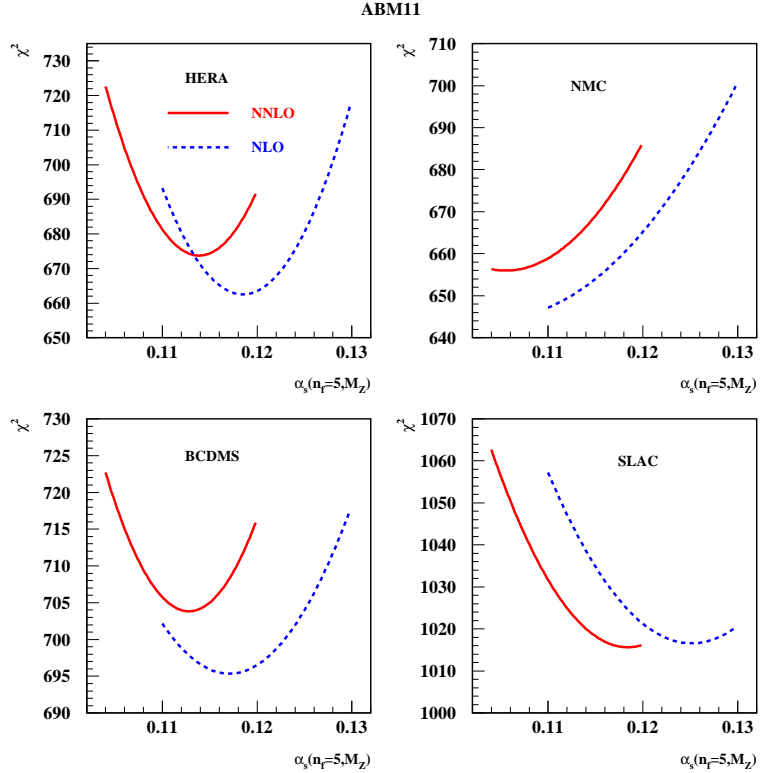,width=4.25in}}
\vspace*{8pt}
  \caption[]{
    The $\chi^2$-profile versus the value of $\alpha_s(M_Z^2)$,
    for the separate data subsets, all obtained in variants of 
    the ABM11 analysis with the value of $\alpha_s$ fixed and 
    all other parameters fitted (solid lines: NNLO fit, dashes: NLO fit); 
    from Ref.~\protect\cite{Alekhin:2012ig}.}
  \protect\label{fig:scans}
\end{figure}
The main version of our analysis is performed at NNLO using  
the three-loop anomalous dimensions in the PDF evolution 
and corresponding Wilson coefficients 
for the light-flavor DIS structure functions and the Drell-Yan process. 
For the neutral-current (NC) heavy-quark contribution we employ the 
approximate NNLO Wilson coefficients~\cite{Kawamura:2012cr}. 
These terms were derived combining results obtained with the soft-gluon resummation 
technique and the high-energy limit of the DIS structure
functions~\cite{Catani:1990eg}. These two 
approaches provide a good approximation 
at kinematics close to threshold of the heavy-quark production
and far beyond the threshold, respectively. Between these two 
regimes the constraint coming from the available NNLO 
massive operator-matrix-element (OME) Mellin moments~\cite{Bierenbaum:2009mv} 
are employed. 
A remaining uncertainty in the NNLO Wilson coefficient obtained in this way is 
quantified by its margins, A and B. To find the best shape of the NNLO term 
preferred by the data we use a linear interpolation between 
these margins
\begin{equation}
  c_{2}^{\,(2)} \,=\,
  (1-d_N) c_{2}^{\,(2),A} + d_N c_{2}^{\,(2),B}
  \, .
\label{eq:inter}
\end{equation}
and fit the interpolation parameter $d_N$ to the data simultaneously with the 
PDF parameters, $\alpha_s$ and $m_c$. 
The charged-current (CC) heavy-quark production is calculated with account 
of the NLO corrections~\cite{Gottschalk:1980rv,Gluck:1996ve,Blumlein:2011zu},
which are the highest-order ones presently available.
Both NC and CC massive Wilson coefficients used in our analysis are
used with the $\overline{\rm MS}$-definition of the heavy-quark mass. 
If compared to the case using the pole-mass the present choice is perturbatively
more stable \cite{Alekhin:2010sv}. With the running-mass definition this contribution 
basically vanishes since the mass can be defined at the typical renormalization/factorization 
scale of the process considered. 

The leading-twist terms provided by the QCD-improved parton model 
are not sufficient at small $Q^2$ and/or final-state hadronic mass $W$, where 
parton correlations cannot be neglected. To account for these
we add on the top of the leading-twist term the 
twist-4 contribution to the DIS structure functions $F_{2,T}$, 
parameterized in a  model-independent form using
spline interpolation~\footnote{The twist-6 terms were also checked in the 
fit and found comparable to zero within errors
applying the cut of $Q^2 > 2.5~{\rm GeV}^2$ used in the present analysis.}. 
The twist-4 spline 
coefficients are fitted to the data together with other the parameters. 
This is particularly important for the case of $\alpha_s$ in view of 
its strong correlation with the higher twist terms. 

\section{Strong Coupling Constant}
\label{sec:alphas}

%
\begin{figure}[pt]
\centerline{\psfig{file=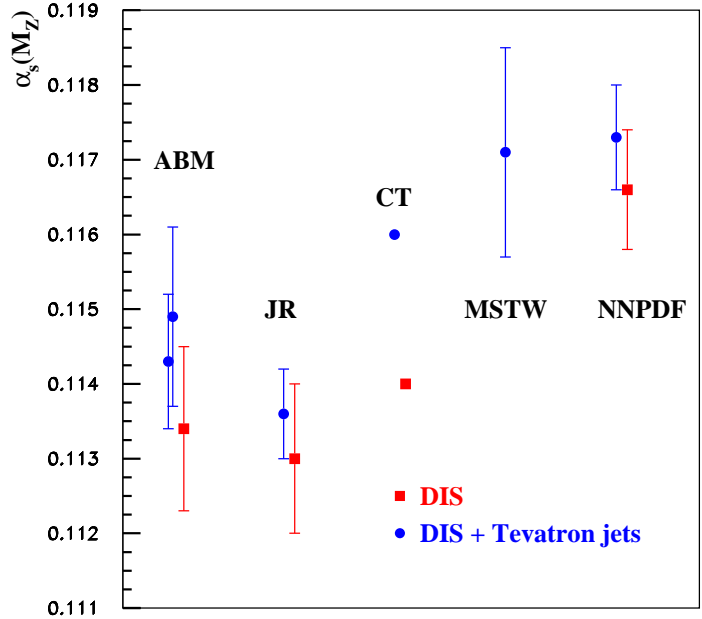,width=3.75in}}
\vspace*{8pt}
  \caption{The values of $\alpha_s(M_Z^2)$ obtained in the ABM 
   fit~\protect\cite{Alekhin:2012ig} in comparison to the ones obtained 
   by JR~\protect\cite{JRnew}, CTEQ~\protect\cite{Gao:2013xoa}, 
   MSTW~\protect\cite{Martin:2009bu}, and 
   NNPDF~\protect\cite{Ball:2011us}  
   from the analysis of the DIS data (squares) 
   and from the combination of the DIS and  
   jet Tevatron data~\protect\cite{Abulencia:2007ez,Abazov:2008hua} 
   (circles). 
}
  \protect\label{fig:comp}
\end{figure}
The strong coupling constant $\alpha_s$ can be determined  
by comparing the $Q^2$-dependence of 
the DIS cross section measurements with the predictions based on the 
QCD-improved paton model. In our analysis the value of $\alpha_s$ is 
obtained simultaneously with the nucleon PDFs and the twist-4 terms. 
This allows to take into account the correlation of $\alpha_s$ with 
other parameters affecting the data $Q^2$-dependence. 
The central value of $\alpha_s$ obtained in 
this way depends on the perturbative order and reduces from NLO to NNLO.
In particular, the ABM11 fit~\cite{Alekhin:2012ig} yields 
\begin{eqnarray}
  \label{eq:alphas-nlo}
  \alpha_s(M_Z^2) \,\,=&
  0.1180\, \pm 0.0012 (\text{exp})\,
  \hspace*{30mm}
  &{\rm NLO} \nonumber
  \, ,
  \\
  \label{eq:alphas}
  \alpha_s(M_Z^2) \,\,=&
  0.1134\, \pm 0.0011 (\text{exp})\,
  \hspace*{30mm}
  &{\rm NNLO}
  \, .
\end{eqnarray}
The values of $\alpha_s$ preferred by each particular DIS data set are 
demonstrated in Fig.~\ref{fig:scans} by means of the $\chi^2$-profiles
obtained in the variants of the ABM11 fit with $\alpha_s$ fixed 
at the values in the range of $0.104\div0.130$. The HERA and BCDMS 
data sets have a similar $\chi^2$-shape with minima around the 
values Eq.~(\ref{eq:alphas}), while the SLAC and NMC data pull 
the value of $\alpha_s$ somewhat up and down, respectively. 
Note that the two latter sets
are sensitive to higher-twist terms due to 
substantial small-$Q^2$ contributions  in these samples. In contrast, 
the HERA and BCDMS data are far less sensitive to higher twists terms
and it is worth noting that in the variant of 
the ABM11 fit excluding the SLAC and NMC data and setting the higher twist 
terms to zero we find $\alpha_s(M_Z^2)=0.1133\, \pm 0.0011 (\text{exp.})$ at NNLO.
The good agreement of this value with Eq.~(\ref{eq:alphas})  
substantiates the consistency between different data sets in our analysis
once the higher twist terms are taken into account. Moreover, this 
cross-check confirms that the combination of the BCDMS and HERA data 
can be used for accurate determination of $\alpha_s(M_Z^2)$ since these two data
sets provide complimentary constraints on the PDFs~\cite{Adloff:2000qk},
see also~\cite{Alekhin:2013kla}.

The NNLO value of $\alpha_s$ Eq.~(\ref{eq:alphas}) is in a good agreement 
with the results of the JR analysis~\cite{JRnew} and the recent 
CTEQ determination~\cite{Gao:2013xoa}, 
while the MSTW~\cite{Martin:2009bu} and NNPDF~\cite{Ball:2011us} 
groups report substantially bigger values,
cf. Fig.~\ref{fig:comp}. 
The discrepancy with MSTW can be explained in part by impact 
of the jet Tevatron data, which pull the value of $\alpha_s(M_Z^2)$ up 
by $0.001\div 0.002$, depending on the fit details.
Furthermore, changing our fit ansatz in direction of the MSTW and NNPDF ones we 
approach their value of $\alpha_s$. In particular, 
dropping the higher twist terms simultaneously 
with an additional cut of $W^2>12.5~{\rm GeV}^2$
imposed by MSTW and NNPDF we
obtain $\alpha_s(M_Z^2)=0.1191\, \pm 0.0006 (\text{exp.})$. In a similar way, 
disregarding the error correlation in the HERA and NMC data, like 
in the MSTW analysis, we obtain $\alpha_s(M_Z^2)$ shifted by 
$+0.0026$, in the direction of the MSTW value. 
Note that in this context the recently updated JR analysis~\cite{JRnew} 
treats the  higher twist properly. 
The CTEQ analysis~\cite{Gao:2013xoa} seems to be  
less sensitive to the impact of the higher twist 
contributions, compared to that by MSTW and NNPDF due to a more stringent cut on 
$Q^2$. 

\section{The Mass of the Charm Quark}
\label{sec:mass}
%
\begin{figure}[pt]
\centerline{\psfig{file=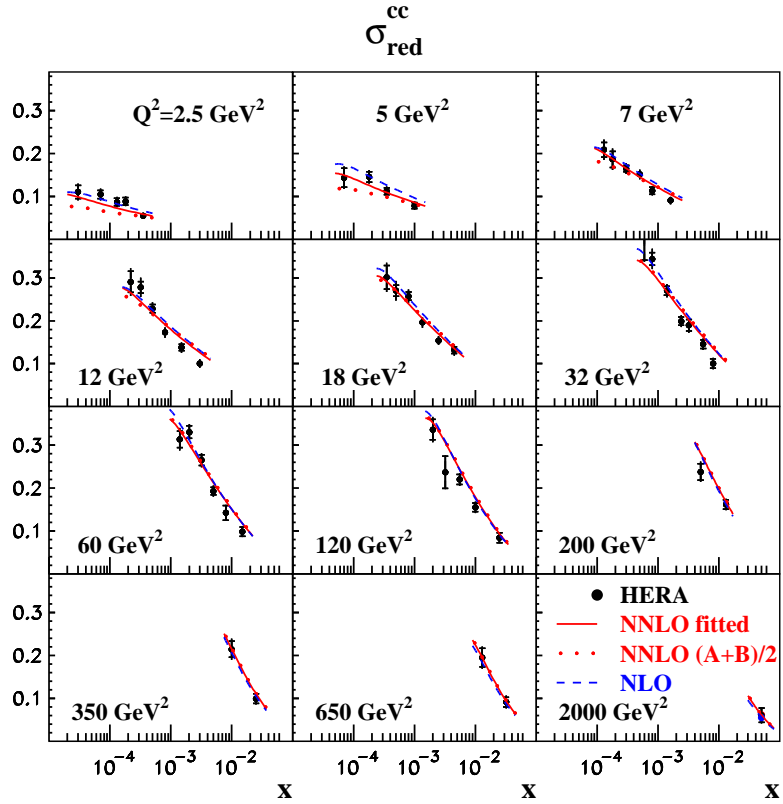,width=4.25in}}
\vspace*{8pt}
\caption{\protect\label{fig:herac}                                              
        The combined HERA data on the reduced cross section for the open 
  charm production~\protect\cite{Abramowicz:1900rp} versus $x$ at different
  values of $Q^2$ in comparison with the analysis of~\protect\cite{Alekhin:2012vu} at   
  NLO (dashed line) and NNLO (solid line) together with 
  a fit variant based on the option (A+B)/2 of the NNLO Wilson
  coefficients of Ref.~\protect\cite{Kawamura:2012cr}, 
  cf. Eq.~(\protect\ref{eq:inter}) (dotted line);
  from Ref.~\protect\cite{Alekhin:2012vu}.
} 
\end{figure}
The sensitivity to the charm quark mass $m_c$ in our analysis appears 
essentially due to the data on 
the NC and CC inclusive DIS~\cite{Aaron:2009aa}, and 
the CC semi-inclusive charm production in DIS~\cite{Bazarko:1994tt,Goncharov:2001qe}
with the most essential experimental constraint on $m_c$ 
coming from the semi-inclusive charm-production 
HERA data~\cite{Abramowicz:1900rp}.
The latter sample comprises the statistics of the H1 and ZEUS experiments
obtained for different $c$-quark decay channels. The combination was 
performed similarly to the case of the inclusive HERA data~\cite{Aaron:2009aa} 
and allows to reduce the systematic errors of each experiment due to cross-calibration 
of the experiments. The FFN scheme provides a good description of the 
semi-inclusive HERA data up to the largest $Q^2$-values covered, 
cf. Fig.~\ref{fig:herac}, with a value of $\chi^2/NDP=61/52$
at NNLO. Here  NDP denotes the
number of data points.    
The $\overline{\rm MS}$-values of $m_c$ found in the analysis of~\cite{Alekhin:2012vu} are
\begin{eqnarray}
  \label{eq:mcres-nlo}
  m_c(m_c) \,\,=&
  1.15\, \pm 0.04 (\text{exp})\,^{+0.04}_{-0.00} (\text{scale})
  \hspace*{30mm}
  &{\rm NLO} 
  \, ,
  \\
  \label{eq:mcres-nnlo}
  m_c(m_c) \,\,=&
  1.24\, \pm 0.03 (\text{exp})\,^{+0.03}_{-0.02} (\text{scale})\,^{+0.00}_{-0.07} (\text{th}),
  \hspace*{14mm}
  &{\rm NNLO_\text{approx}}
  \, ,
\end{eqnarray}
\begin{figure}[pt]
\centerline{\psfig{file=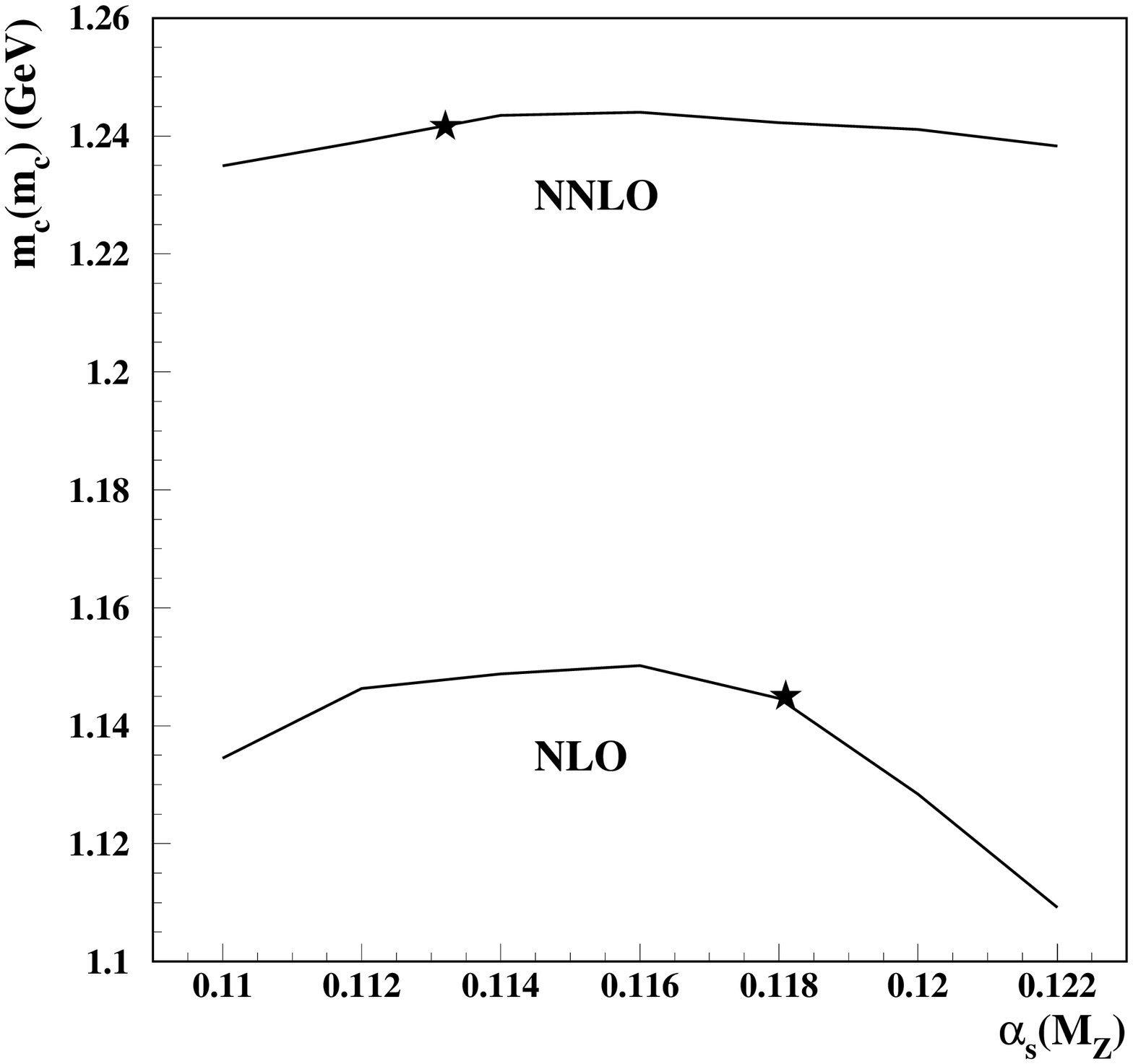,width=2.75in}}
\vspace*{8pt}
\caption{\protect\label{fig:alpha}
  The values of $m_c(m_c)$ obtained in the NLO and NNLO variants of
  our analysis with the value of $\alpha_s(M_Z^2)$ fixed. 
  The position of the star displays the result with the value of
  $\alpha_s(M_Z^2)$ fitted~\protect\cite{Alekhin:2012ig}; 
  from Ref.~\protect\cite{Alekhin:2012vu}.
}
\end{figure}
at NLO and NNLO, respectively, see also~\cite{Alekhin:2012un}.
The experimental accuracy of 30~MeV obtained at NNLO is quite competitive with other 
determinations of $m_c$ based on the $e^+e^-$ data and the central value 
Eq.~(\ref{eq:mcres-nnlo}) is in a good agreement with the world 
average~\cite{Beringer:1900zz}.
The scale error in Eqs.~(\ref{eq:mcres-nlo}) and (\ref{eq:mcres-nnlo}) is obtained
varying the factorization scale by a factor of $1/2$ and $2$ around 
the nominal value of $\sqrt{m_c^2+\kappa Q^2}$, 
where $\kappa=4$ for NC and $\kappa=1$ for  CC
heavy-quark production, respectively. 
In the NNLO case an additional error related to the uncertainty 
in the massive Wilson coefficients contributes. The value of 
Eq.~(\ref{eq:mcres-nnlo}) is obtained for the interpolation parameter 
$d_N=-0.1$ being preferred by the fit, roughly corresponding to option A 
of the Wilson coefficients of Ref.~\cite{Kawamura:2012cr}. Meanwhile, 
option B is clearly excluded by the data with $\chi^2$/NDP=115/52. 
Therefore the uncertainty due to the missing NNLO massive terms 
is estimated as a variation between options A and (A+B)/2 which yields
the value of 70~MeV Eq.~(\ref{eq:mcres-nnlo}). 
The value of $m_c(m_c)$ obtained in our analysis demonstrates remarkable 
stability w.r.t. $\alpha_s(M_Z^2)$. 
Performing variants of our analysis with the values of $\alpha_s$
fixed in the wide range around the best value preferred by the 
data, we find a variation of $m_c(m_c)$ in the range of 10-20~MeV, 
depending on the order, cf. Fig~\ref{fig:alpha}.

The NLO value of $m_c(m_c)$ obtained in our analysis is somewhat lower 
than the one of $m_c(m_c)=1.26\pm0.05~(\text{exp})~{\rm GeV}$ from 
the analysis based on the HERA data only~\cite{Abramowicz:1900rp}. To
understand the difference we checked the cases of a cut on 
$Q^2=3.5~{\rm GeV}^2$, likewise in the HERA fit, and no 
semi-inclusive Tevatron data~\cite{Bazarko:1994tt,Goncharov:2001qe} 
included. As a result we obtain 
shifts in $m_c(m_c)$ by $+30~\text{MeV}$ and $+40~\text{MeV}$, respectively.
The remaining discrepancy should be attributed to the particularities
of the HERA PDFs. The value of $m_c(m_c)$ was recently also determined by 
the CTEQ collaboration~\cite{Gao:2013wwa}. In contrast to our case this 
determination is based on the S-ACOT-$\chi$ 
prescription as GMVFN scheme. Furthermore, the $\overline{\rm MS}$ 
coefficients of Ref.~\cite{Gao:2013wwa} are obtained by straightforward 
substitution of the pole- and running-mass matching relation 
into the pole-mass coefficients. The expressions obtained in such a way 
correspond to a mixed order in $\alpha_s$. Moreover, the advantage 
of this approach is not evident in view of the poor perturbative convergence 
of the mass matching relation. The central CTEQ value of 
$m_c(m_c)=1.12 ^{+0.11}_{-0.17}~{\rm GeV}$ is lower than  
the world average, while the CTEQ errors are much larger than those in 
Eqs.~(\ref{eq:mcres-nlo}) and (\ref{eq:mcres-nnlo}) due to 
the impact of the uncertainty in the GMVFN scheme modeling. 
%
\begin{figure}[pt]
\centerline{\psfig{file=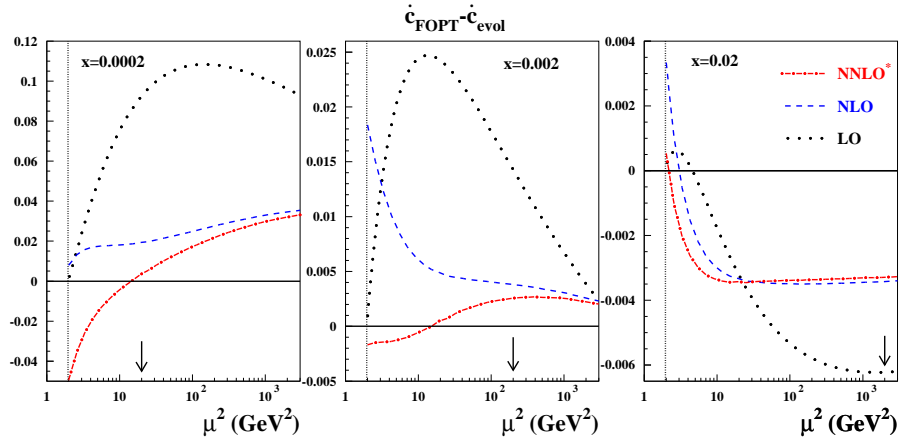,width=4.75in}}
\vspace*{8pt}
\caption{\label{fig:pdfder} The difference between the 
$c$-quark PDFs derivatives 
$\dot{c}(x,\mu^2)\equiv \frac{dc(x,\mu^2)}{d\ln\mu^2}$ calculated with 
the FOPT matching condition and with the massless 4-flavor evolution 
starting at the matching point $\mu_0=m_c=1.4~{\rm GeV}$ versus the
factorization scale $\mu^2$ at different values of $x$ in the LO, 
NLO, and NNLO* approximations. The arrows display upper margin of the 
HERA collider kinematics with the collision 
c.m.s. energy squared $s=10^5~{\rm GeV}^2$ and the vertical lines correspond to
the matching point position $\mu_0$.}
\end{figure}

\section{VFN Uncertainties}
\label{sec:vfn}
The choice of the factorization scheme plays an essential role in the analysis of 
existing DIS data due to important constraints coming from the small-$x$ region, 
where the heavy quark contribution is numerically large. While 
our analysis is based on the FFN scheme, many other groups employ 
different variants of the GMVFN scheme, which differ by {\it modeling} of the 
low-$Q^2$ region. The spread between these variants
is rather substantial and thus implies a corresponding uncertainty in the basic parameters 
determined in these GMVFN fits. In particular, the value of $m_c$ determined from a
combination of the inclusive and semi-inclusive HERA data with the 
different versions of the 
ACOT and RT prescriptions for the VFN scheme demonstrate a 
spread of 400 MeV~\cite{Abramowicz:1900rp} . 
There are also sources of the VFN scheme uncertainties, which are  
common for all these prescriptions. 
Firstly, the matching of heavy-quark PDFs is commonly performed
at the factorization scale $\mu$ equal to the 
$\mu_0=m_{c}$, resp. $m_b$. Clearly, at these scales neither of the heavy flavors can 
be dealt with as massless. The matching point $\mu_0$
is not fixed by theory 
and in principle it can vary in a wide range being an artefact of the description
not contributing to the observables according to the renormalization group equations. 
Additional 
uncertainty emerge for the 4(5)-flavor PDFs 
in the NNLO analysis. They are commonly matched with the 
3(4)-flavor ones using {\it NLO matching condition}, 
since the NNLO OMEs are not yet known in the complete form~\footnote{
For progress in this field 
cf.~\cite{Ablinger:2012ej,Ablinger:2012qm,Ablinger:2010ty,Blumlein:2012vq}.}.
However, to provide consistency with the NNLO Wilson coefficients  
the evolution of these PDFs is performed in the NNLO approximation that
introduces an additional uncertainty due to missing higher-order
corrections into the analysis.
At the same time, the evolution of the 4(5)-flavor PDFs in the VFN scheme
leads to a resummation of the terms $\sim \ln(\mu^2)$ 
which in part reproduce the higher-order corrections, being known, however, not to be dominant.
Relations between the resummation effects and the VFN evolution 
uncertainties are illustrated in Fig.~\ref{fig:pdfder} by comparison 
of the $\mu$-derivatives for the $c$-quark distribution 
being calculated in different ways. In one case the distributions 
are matched at the scale of $\mu_0$ using the fixed-order-perturbation-theory
(FOPT) matching conditions and then evolved starting from $\mu_0$ with 
the massless splitting functions. In another case they are calculated 
with the FOPT matching conditions at all scales. 
The difference between these two cases do not demonstrate a significant 
rise with $\mu$.  The only exclusion is observed at $x\lesssim 0.0001$
and at scales outside of the kinematics being probed in experiment. 
Therefore it cannot be attributed to the impact of the log-term resummation. 
In contrast, there is a substantial difference in the derivatives 
calculated in the NLO and NNLO*, i.e. the combination of the NLO 
matching with the NNLO evolution. This difference yields an estimate of 
the uncertainty in the VFN scheme due to the missing higher-orders, which 
is obviously larger than the resummation effects. 
Checking the impact of this uncertainty w.r.t. the value of $\alpha_s(M_Z^2)$
in combination with the variation of the matching 
point for the 4-flavor PDFs in the range of $1.2\div 1.5~{\rm GeV}$
we find a value of $\pm 0.001 $. This is comparable to the experimental uncertainty 
and makes the VFN schemes incompetitive with the FFN one in the 
precision determination of $\alpha_s(M_Z^2)$. 

\section*{Acknowledgments}

We thank P. Jimenez-Delgado and E. Reya for discussions. This work has been supported in part 
by Helmholtz Gemeinschaft under contract VH-HA-101 ({\it Alliance Physics at the Terascale}), 
DFG Sonderforschungsbereich/Transregio~9 and by the European Commission through contract 
PITN-GA-2010-264564 ({\it LHCPhenoNet}).


\begin{thebibliography}{0}

\bibitem{Bethke:2011tr}
  S.~Bethke {\it et al.},
  {\sf Workshop on Precision Measurements of $\alpha_s$},
  arXiv:1110.0016 [hep-ph].

\bibitem{Kawamura:2012cr} 
  H.~Kawamura, N.~A.~Lo Presti, S.~Moch and A.~Vogt,
  {\it Nucl.\ Phys.\ B} {\bf 864}, 399 (2012).

\bibitem{Alekhin:2010sv} 
  S.~Alekhin and S.~Moch,
  {\it Phys.\ Lett.\ B} {\bf 699}, 345 (2011).

\bibitem{Alekhin:2012ig} 
  S.~Alekhin, J.~Bl\"{u}mlein and S.~Moch,
  {\it Phys.\ Rev.\ D} {\bf 86}, 054009 (2012);

\bibitem{Catani:1990eg} 
  S.~Catani, M.~Ciafaloni and F.~Hautmann,
  {\it Nucl.\ Phys.\ B} {\bf 366}, 135 (1991).

\bibitem{Bierenbaum:2009mv} 
  I.~Bierenbaum, J.~Bl\"{u}mlein and S.~Klein,
  {\it Nucl.\ Phys.\ B} {\bf 820}, 417 (2009).

\bibitem{Gottschalk:1980rv} 
  T.~Gottschalk,
  {\it Phys.\ Rev.\ D} {\bf 23}, 56 (1981).

\bibitem{Gluck:1996ve} 
  M.~Gl\"uck, S.~Kretzer and E.~Reya,
  {\it Phys.\ Lett.\ B} {\bf 380}, 171 (1996)
  [Erratum-ibid.\ B {\bf 405}, 391 (1997)].

\bibitem{Blumlein:2011zu}
  J.~Bl\"umlein, A.~Hasselhuhn, P.~Kovacikova and S.~Moch,
  {\it Phys.\ Lett.\ B} {\bf 700} 294 (2011).

\bibitem{Adloff:2000qk} 
  C.~Adloff {\it et al.}  [H1 Collaboration],
  {\it Eur.\ Phys.\ J.\ C} {\bf 21}, 33 (2001).

\bibitem{Alekhin:2013kla}
  S.~Alekhin, J.~Bl\"{u}mlein and S.~Moch,
  arXiv:1303.1073 [hep-ph].

\bibitem{JRnew} 
  P.~Jimenez-Delgado and E.~Reya, private communication. 

\bibitem{Gao:2013xoa} 
  J.~Gao {\it et al.},
  arXiv:1302.6246 [hep-ph].

\bibitem{Martin:2009bu} 
  A.~Martin, W.~J.~Stirling, R.~Thorne and G.~Watt,
  {\it Eur.\ Phys.\ J.\ C} {\bf 64}, 653 (2009).

\bibitem{Ball:2011us} 
  R.~D.~Ball {\it et al.},
  {\it Phys.\ Lett.\ B} {\bf 707}, 66 (2012).

\bibitem{Abulencia:2007ez}
A.~Abulencia, et~al., {\it Phys.Rev. D} {\bf 75}, 092006 (2007).

\bibitem{Abazov:2008hua}
V.~Abazov, et~al., {\it Phys.Rev.Lett.} {\bf 101}, 062001 (2008).

\bibitem{Abramowicz:1900rp} 
  H.~Abramowicz {\it et al.}  [H1 and ZEUS Collaborations],
  {\it Eur.\ Phys.\ J.\ C} {\bf 73}, 2311 (2013).

\bibitem{Aaron:2009aa} 
  F.~D.~Aaron {\it et al.}  [H1 and ZEUS Collaboration],
  {\it JHEP {\bf 1001}, 109} (2010).

\bibitem{Bazarko:1994tt} 
  A.~O.~Bazarko {\it et al.}  [CCFR Collaboration],
  {\it Z.\ Phys.\ C} {\bf 65}, 189 (1995).

\bibitem{Goncharov:2001qe} 
  M.~Goncharov {\it et al.}  [NuTeV Collaboration],
  {\it Phys.\ Rev.\ D} {\bf 64}, 112006 (2001).

\bibitem{Alekhin:2012vu}
  S.~Alekhin {\it et al.},
  {\it Phys.\ Lett.\ B} {\bf 720} (2013) 172

\bibitem{Alekhin:2012un}
  S.~Alekhin, K.~Daum, K.~Lipka and S.~Moch,
  {\it Phys.\ Lett.\ B} {\bf 718} (2012) 550

\bibitem{Beringer:1900zz} 
  J.~Beringer {\it et al.}  [Particle Data Group Collaboration],
  {\it Phys.\ Rev.\ D} {\bf 86}, 010001 (2012).

\bibitem{Gao:2013wwa} 
  J.~Gao, M.~Guzzi and P.~M.~Nadolsky,
  arXiv:1304.3494 [hep-ph].

\bibitem{Ablinger:2012ej}
  J.~Ablinger {\it et al.},
  {\it PoS LL} {\bf 2012} (2012) 033.

\bibitem{Ablinger:2012qm}
  J.~Ablinger {\it et al.},
  {\it Nucl.\ Phys.\ B} {\bf 864} (2012) 52.

\bibitem{Ablinger:2010ty}
  J.~Ablinger {\it et al.},
  {\it Nucl.\ Phys.\ B} {\bf 844} (2011) 26.

\bibitem{Blumlein:2012vq}
  J.~Bl\"umlein, A.~Hasselhuhn, S.~Klein and C.~Schneider,
  {\it Nucl.\ Phys.\ B} {\bf 866} (2013) 196.



\end{thebibliography}
\end{document}